\documentclass[letter]{aa} 

\usepackage{natbib}
\bibpunct{(}{)}{,}{a}{}{,} 
\usepackage{amsmath}
\usepackage{graphicx}
\usepackage{longtable}
\usepackage{txfonts}
\usepackage{color}
\usepackage{soul}

\newcommand{\ndhp}{N$_2$H$^+$}
\newcommand{\ndhpb}{N$_2$H$^+$ }

\newcommand{\cc}{cm$^{-3}$}

\newcommand{\sqc}{cm$^{-2}$}

\newcommand{\mjy}{MJy\,sr$^{-1}$}
\newcommand{\kjy}{kJy\,sr$^{-1}$}

\newcommand{\Av}{A$_\mathrm{V}$}
\newcommand{\Avb}{A$_\mathrm{V}$ }
\newcommand{\Rv}{R$_\mathrm{V}$}
\newcommand{\Rvb}{R$_\mathrm{V}$ }

\newcommand\mic{$\mu$m}
\newcommand\micb{$\mu$m }
\newcommand{\Ibg}{I$_\mathrm{bg}$ }

\begin{document}
        
        \title{On the importance of scattering at 8\,\mic : Brighter than you think
        }
        
        \author{C. Lef{\`e}vre            \inst{1}
                \and
                L. Pagani\inst{1}
                \and
                M. Min\inst{2}
                \and
                C. Poteet\inst{3}
                \and
                D. Whittet\inst{3}}
        
        \offprints{C.Lef{\`e}vre}
        
        \institute{ LERMA \& UMR8112 du CNRS, Observatoire de
                Paris, 61, Av. de l'Observatoire, 75014 Paris, France
                \and
                Sterrenkundig Instituut Anton Pannekoek, University of Amsterdam, Science Park 904, 1098 XH Amsterdam, The Netherlands
                \and
                Department of Physics, Applied Physics and Astronomy and New York Center for Astrobiology,\\
                Rensselaer Polytechnic Institute, 110 Eighth Street, Troy, NY 12180, USA\\
                \email{charlene.lefevre@obspm.fr}
        }
        
        \date{Received July, 20, 2015; accepted December, 7, 2015}
        

        \abstract
        {Extinction and emission of dust models need for observational constraints to be validated. The coreshine phenomenon has already shown the importance of scattering in the 3 to 5 $\mu$m range and its ability to validate dust properties for dense cores.} 
        {{We want to investigate whether scattering can also play a role at longer wavelengths and to place even tighter constraints on the dust properties.}}
        {We {analyze} the inversion of {the Spitzer 8\,\micb map of the dense molecular cloud L183}, to {examine} the importance of scattering {as a potential contributor to the line-of-sight extinction.}}
        {The column density deduced from the inversion of the 8\,\micb map, when we neglect scattering, disagrees with all the other column density measurements of the same region. Modeling confirms that scattering at 8\,\micb is not negligible with an intensity of several hundred \kjy. {This demonstrates} the need of efficiently scattering dust grains at MIR wavelengths up to 8\,\mic. Coagulated aggregates are good candidates and might also explain the discrepancy at high extinction between E(J--K) and $\tau_{9.7}$ {toward dense molecular clouds. Further investigation requires considering efficiently scattering dust grains including ices as realistic dust models.}}
        {}   
        
        
        \maketitle
        
        
        \section{Introduction}
        
        { Although star formation occurs deep inside molecular cores, the process of conversion from the interstellar medium reservoir into cores, and ultimately into protostars, is still being debated.} The mass distribution inside clouds has consequences {for} their equilibrium and their ability to form prestellar cores. Molecular clouds can be divided into two categories: translucent clouds (with a visible extinction of \Av$\sim$1--5\,mag) and dark molecular clouds (\Av$>$5\,mag). The latter may contain gas and dust density peak(s), the so--called self--gravitating core(s). A common {threshold of extinction} of \Av$\sim$8\,mag \citep{2014Andre}, above which star formation can occur, has been claimed for molecular clouds in the solar vicinity. Nevertheless, this threshold {has not been found for clouds located farther than the Gould Belt} \citep{2014Schisano,2015Montillaud}. {It} stresses that the mass distribution has to be characterized from core to core. 
        
        Cloud and core column density maps can be obtained by different methods: molecular tracers, far--infrared dust emission, or near--infrared (NIR) stellar extinction {\citep{2009Goodman}}. Molecular tracers give high--angular resolution maps, but do not systematically peak at the same position, and can be depleted onto grains. Moreover, their {abundances} rely on excitation conditions and chemical network. Far--infrared dust emission as seen by the {\textit{Herschel} Space Observatory} reveals the cold dust deeply {hidden in molecular clouds. However,} the column density of the coldest grains cannot be retrieved {through} spectral energy distribution fitting, even {when} including {submillimeter} data. Indeed, the presence of surrounding warmer dust can lead to { underestimating} the core mass by a factor of 3 \citep{2015Pagani}. Finally, star counts, as well as NIR stellar extinction, are cloud-model independent, but {may} suffer {from both} the cloud location and density effects. The number of stars decreases by a factor of 10 with {$\sim$3 magnitudes of extinction}, which leads to a typical limit of 40\,mag for clouds above the {Galactic} plane. {Reliable column densities can be obtained by these methods \citep{2008Foster,2009Lombardi}, but despite their accuracy at low and intermediate \Av, they are not adapted to scrutinizing high density cores. Moreover, at high Galactic latitude, the resolution may degrade to typically less than 1'.}\\
        Other studies have chosen to rely on the surface brightness extinction {at mid-infrared wavelengths to estimate column density toward dense regions}  \citep[e.g.,][]{2000Bacmann,2009Stutz,2012Butler}. {The use of the \textit{Spitzer} 8\,\micb map gives the opportunity} to reach a resolution of {2.8\arcsec}, a factor of $\sim$4 better than what can be obtained with dust emission or {molecular lines} with single-dish observations. In principle, by knowing the background intensity behind the cloud, I$_\mathrm{bg}$, and inverting the 8\,\micb map, one can retrieve reliable column densities and masses {in a wide range of visual extinctions up to \Av$\sim$200\,mag}. {Two assumptions are needed:} the 8\,\micb opacities have to be relatively unresponsive to the dust properties, and the scattering at this wavelength has to be negligible. {

The importance of scattering has already been probed in the 3--5\,\micb wavelength range by the detection of coreshine in more than 100 molecular clouds \citep{2010Pagani,2014Lefevre}. Coreshine appears when scattering is strong enough to overpass the background extinction and make the densest regions shine owing to the presence of large grains \citep{2010Steinacker}}. In this letter, we investigate the impact of neglecting {dust scattering} at 8\,\micb {toward} the L183 molecular cloud {and compare our results with existing dust models available in the literature.}
        
        \vspace{-0.5cm}
        \section{Observations and analysis}\label{sect:inversion}
        
        L183 is a molecular cloud located 36.7 degrees above the {Galactic} plane, for which NIR maps at J, H, and K$_\mathrm{s}$ bands were performed with the Visible and Infrared
        Survey Telescope for Astronomy (VISTA). These large scale maps (1.6 square degrees) allow us to complete the previous CFHTIR data from \cite{2004Pagani} with a wider field of view. {To sample the highly extinguished regions, we also included} the \textit{Spitzer}/IRAC observations (R. Paladini, PID: 80053 and C. Lawrence, PID: 94). Using these data, we built three source catalogs to be converted into extinction maps: c1 (J, H, K$_\mathrm{s}$), c2 (H, K$_\mathrm{s}$, 3.6\,\mic), and c3 (3.6\,\mic, 4.5\,\mic). All three catalogs sample different regions and were combined to obtain our reference column density map (C$_\mathrm{ext}$). Thus, the conversion of the surface brightness of the \textit{Spitzer} 8\,\micb map into extinction map (8$_\mathrm{ext}$) can be compared to the C$_\mathrm{ext}$ reference map. Below, we present the technical details of how these maps are constructed.\\
        The source catalogs were {built} {using the SExtractor routine} \citep{1996Bertin}, {and nearly 78000 stars were detected in all three NIR bands and 8500 in both the 3.6 and 4.5\,\micb bands, but on a smaller surface}.  The final image calibration was done by {cross--correlating} with the Two Micron All Sky Survey (2MASS) point source catalog for the NIR range, and the Wide-field Infrared Survey Explorer (WISE) catalog at MIR wavelengths. The completeness magnitudes are down to {20.5 for J, 19.7 for H, 19 for K$ _\mathrm{s}$, 18 at 3.6\,\mic, and 17.5 at 4.5\,\mic}. 

To obtain a robust column density map from the catalogs, we used the constant resolution NICER algorithm \citep{2001Lombardi} with a nearby reference field {from the} 2MASS and WISE catalogs. The NICER algorithm relies on multi--band catalogs to deduce the extinction and reduce the noise toward  {low extinction regions}. We used it with both the J, H, and K$_\mathrm{s}$ catalogs (c1) and the H, K$_\mathrm{s}$ and 3.6\,\micb catalogs (c2). The two maps (c1, c2) were converted to visual extinction maps {assuming} the A$_\lambda$/A$_V$ ratio conversion from dust models. While the NIR conversion ratios for c1 are not very sensitive to the dust model, the conversion {to obtain} c2 strongly depends on the dust model. As a result, the adopted A$_\lambda$/A$_V$ coefficients were taken from the \citet[WD01]{2001Weingartner} \Rv = 3.1 dust model for external regions (c1) and \Rv = 5.5 {(case B including large grains, hereafter 5.5B)} dust model {to obtain} the c2 map. {Indeed, the 5.5B dust model correctly reproduces the extinction law toward dense molecular clouds, even if it does not include ices} \citep{2013Ascenso}. Nevertheless, because of the decreasing number of stars with extinction, the two maps do not go beyond 40 mag, so we also used the statistical information of star distribution from our 3.6/4.5 $\mu$m catalog. 

\begin{figure}
                \centering
                \includegraphics[width=0.75\linewidth]{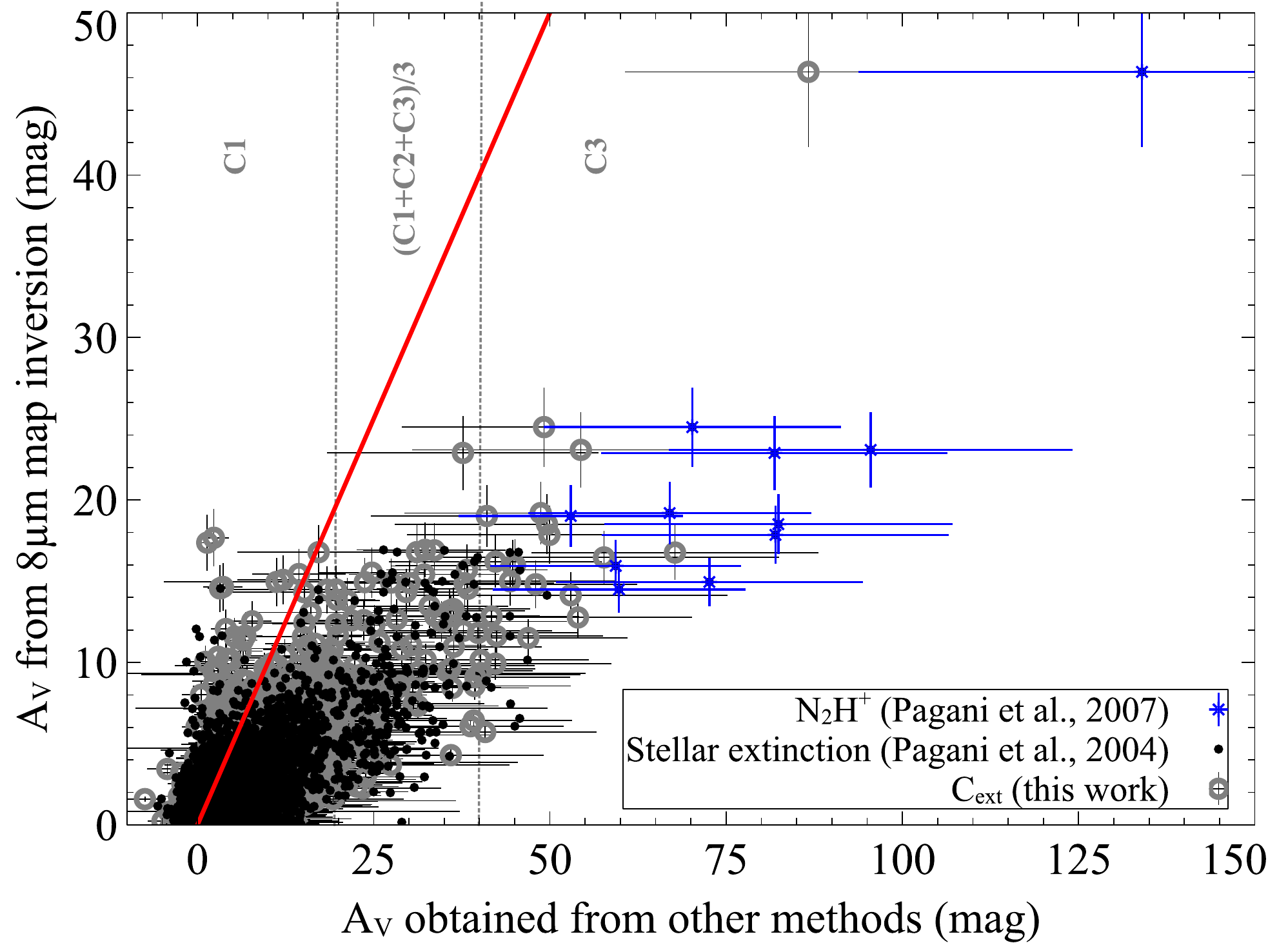}
                \caption{{Comparison between \Avb obtained from different methods at the 20" resolution. The \Avb obtained from the 8\,\micb map inversion is associated to the highest peak value (I$_\mathrm{bg,min}$\,=\,0.42\,\mjy). The red line shows the one-to-one relation. Gray lines and C1, C2, C3 refer to the different catalogs used to build the extinction map.}}
                \label{fig:Av_converted}
                \vspace{-0.5cm}    
        \end{figure}

This star-counting method {provides} a third map (c3) with a better estimate of the extinction toward the dense core in the region where c1 and c2 lack information. The c1, c2, and c3 maps were all smoothed to a resolution of 1' to ensure that enough stars are present in each grid cell. The c1, c2, and c3 maps were finally combined by {retaining} c1 values below 20 mag, the mean {of the three maps} between 20 and 40 mag, and only c3 beyond. {The visual extinctions we obtained (C$_\mathrm{ext}$) are recorded on the x axis of Fig. \ref{fig:Av_converted}}.
         
        The first step in obtaining 8$_\mathrm{ext}$ was to subtract the stellar and background intensities from \textit{Spitzer} maps, following the method of \cite{2014Lefevre}. {The residual signal} ($\Delta$) {decreases} to $-350$\,\kjy($\Delta_\mathrm{min}$) toward the prestellar core (PSC) at 8\,\mic. If we suppose that the extinction is only due to absorption by the dust,  as in \cite{2000Bacmann}, it is directly related to $\Delta$ by $\Delta$\,=\,\Ibg${(\mathrm{e}^{-\tau_8}-1)}$, where $\tau_8$ is the 8\,\micb dust opacity integrated {along} the line of sight, and \Ibg {is} the background intensity value. {Using nearby stars for which the stellar type is known (\citealt{2013Whittet} and Fig. {\ref{fig:spatial_discr}}), the value of \Ibg was independently estimated from the 8\,\micb map.} Among the eight stars falling onto the 8\,\micb map, we kept the ones for which the $\Delta$ value is reliable (3$\sigma$\,$\sim$\,60\,\kjy). Since we make the approximation that the $\tau_8$ {value from each star is locally
                the same as the one in its vicinity}, we also eliminated the H star, saturated at 8\,\mic. We were left with three stars (B, D, I) with a known E(J--K)=(J-K)$_\mathrm{2MASS}-$(J-K)$_0$. From their E(J--K), we obtained the $\tau_8$ values by assuming {A$_8$/A$_\mathrm{K}$} conversion coefficients. We adopted the same conversion coefficients as the ones used to obtain the c1 and c2 maps, which give a range of possible values for $\tau_8$ (Table \ref{tab:stars}).
                       \vspace{-0.5cm}         
        \begin{table}[h]
                \caption{Background values calculated at the vicinity of the stars. {The star names and E(J--K) values} refer to \cite{2013Whittet} and the (3.1) and (5.5B) notations to the dust models from \cite{2001Weingartner}.}
                \label{tab:stars}
                \begin{tabular}{ccccccc}
                        \hline
                        Star & E(J--K) & $\tau_\mathrm{8}$ & $\tau_\mathrm{8}$& $\Delta$ & I$_\mathrm{bg, min}$ & I$_\mathrm{bg, max}$\\ 
                        & mag & 3.1& 5.5B& \kjy & \mjy & \mjy \\
                        \hline
                        B & 0.62 & 0.067& 0.15 & --64 $\pm$ 6 & 0.42 & 1.08 \\ 
                        D  & 0.70 & 0.076& 0.17 & --62 $\pm$ 6 & 0.36 & 0.93 \\                              
                        I & 2.49 & 0.27 & 0.60 & -80 $\pm$ 8 & 0.16 & 0.37 \\ 
                        \hline 
                \end{tabular} 
        \end{table}
        
        \noindent The calculated background values are minimal (I$_\mathrm{bg, min}$) for WD01(5.5B) and maximal (I$_\mathrm{bg, max}$) for WD01(3.1). The compatible \Ibg values {between stars B and D} range from 0.42 to 0.93\,\mjy\,(Table \ref{tab:stars}), which is comparable to the values proposed by \citet{2014Lefevre} from a different method. Nevertheless, \Ibg calculated from the I star is unreasonably low, since only I$_\mathrm{bg}\geq|\Delta_\mathrm{min}|$ are physically meaningful. {The value of I$_\mathrm{bg, max}$  from Star I is} slightly compatible with the I$_\mathrm{bg, min}$ obtained for {Stars B and D}, but is calculated with a lower \Rvb value. This is not expected because {Star I} is more attenuated than {Stars B and D}, from their E(J--K) values. Given this incompatibility, we rely only on {Stars B and D} for the \Ibg value {and will revisit Star I  in the discussion}. From the \Ibg range, we retrieve the $\tau_8$ maps and adopt {their} associated conversion coefficients to obtain the extinction maps:  A$_8$/A$_\mathrm{V}$(3.1)=0.02 for I$_\mathrm{bg}$=0.93\,\mjy\, and A$_8$/A$_\mathrm{V}$(5.5B)=0.045 for I$_\mathrm{bg}$=0.42\,\mjy.
        
        The peak value obtained for the 8$_\mathrm{ext}$ map (34 or 82 mag) at 2.8\arcsec\, resolution  is too low {when compared to the value obtained from \ndhpb observations}  \citep{2007Pagani,2015Lique}. In fact, {inside a 10\arcsec\,radius, the column density of {N(H$_2$) is 1.2$^{+0.5}_{-0.3}\times$10$^{23}$}\,\sqc}, {which corresponds to an extinction of at least {\Av\,=\,130$_{-32}^{+54}$\,mag} when assuming the \cite{1978Bohlin} conversion.} Moreover, 8$_\mathrm{ext}$ is systematically lower than C$_\mathrm{ext}$ beyond 25 magnitudes of extinction (Fig. \ref{fig:Av_converted}). {The region where the 8\,\micb inversion fails to reproduce stellar extinction corresponds to the northern filament and its surroundings (Fig. \ref{fig:spatial_discr}).} In this region, the low values given by the inversion of the  8\,\micb map could not explain {the diminishing numbers of background stars} {in NIR+MIR}. {

Thanks to the ample number of star counts and {overlapping of the techniques between \Avb ranges}, we have confidence in the \Avb values estimated from {stellar extinction}.} This result is confirmed by the compatibility of our C$_\mathrm{ext}$ map with \ndhpb estimates, as well as dust seen in emission at 1.2\,mm \citep[Fig. \ref{fig:Av_converted}]{2004Pagani}. {Considering more sophisticated methods of building C$_\mathrm{ext}$ would help to reduce the uncertainties (i.e., \citealt{2008Foster}) but would not compensate for the difference between C$_\mathrm{ext}$ and 8$_\mathrm{ext}$ (Fig. \ref{fig:spatial_discr})}. Thus, the only way to explain {the} discrepancy between the 8$_\mathrm{ext}$ map and other column density estimates is by considering that the apparent extinction is weaker than the true extinction owing to an additional component. Polycyclic Aromatic Hydrocarbons (PAHs) have already been excluded in this cloud \citep{2010Steinacker}, {and given the extremely cold temperatures inside the cloud,} we are left with scattering as the {only plausible source} of compensation. 
        
        \vspace{-0.5cm}
        
        \section{Modeling with scattering}\label{sect:back}
        
        If significant, the scattering contribution, I$_\mathrm{sca}$, is part of the measured $\Delta$ value: $\Delta\,=\,$I$_\mathrm{bg}{(\mathrm{e}^{-\tau_8}-1)}$ + I$_\mathrm{sca}$. We estimated I$_\mathrm{sca}$ following several steps. First, we build a representative cloud density model based {on C$_\mathrm{ext}$} and \ndhpb information. To compare the modeling with observations, we adopt a background value {consistent with} the extinction toward Stars B and D at 8\,\micb (Table \ref{tab:stars}). Then, we vary the dust properties inside the cloud density model until finding a compatible solution with the 8\,\micb map. Finally, we validate {our} solution {using} several tests.
        
        
        \begin{figure}[t!]
                \centering
                \includegraphics[width=0.75\linewidth]{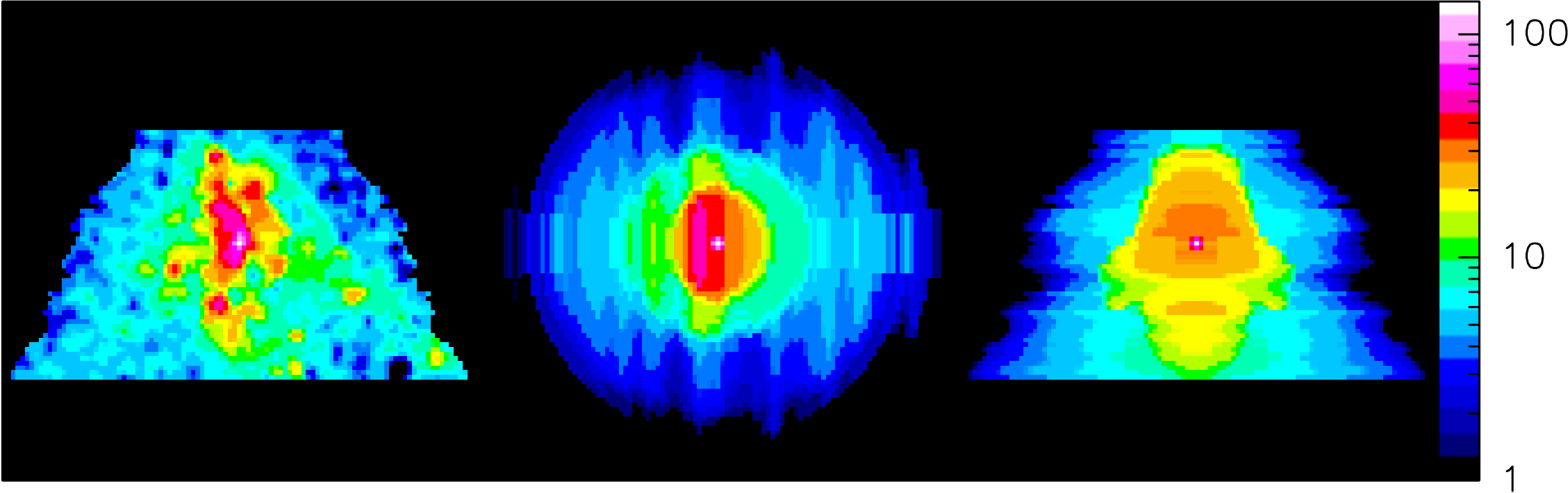}
                \caption{\Avb maps of the cloud model with the common color scale in magnitudes. Left: facing the cloud; middle: from above; right: from the side.}
                \label{fig:cloud_model}
                 \vspace{-0.5cm}    
        \end{figure}
        
        {To study the scattering contribution, it is mandatory to consider the anisotropy of the radiation field, by using a 3D radiative transfer code and take that opportunity to include a 3D cloud model} \citep{2014Lefevre}. First, we use the C$_\mathrm{ext}$ map, and make a {rotational} symmetry to obtain the 3D model. Given the velocity gradient inside the {northern filament} \citep{2009Pagani} and the triangular flaring shape of the cloud, we required that the model be {as elongated in depth as in width}. This first approximation might not be correct, but is sufficient for our needs. A central core compatible with the \ndhpb density profile \citep{2007Pagani, 2015Lique} was placed at the cloud center, and densities were scaled in the third dimension to reproduce the absolute column density (Fig. \ref{fig:cloud_model} -- left).  The resolution adopted for the modeling is 20\arcsec\,per cell to correctly sample the NIR extinction maps.
        
\begin{figure}[h!]
                \centering
                \includegraphics[width=0.55\linewidth, trim=2.5cm 0.5cm 3cm 1cm \clip]{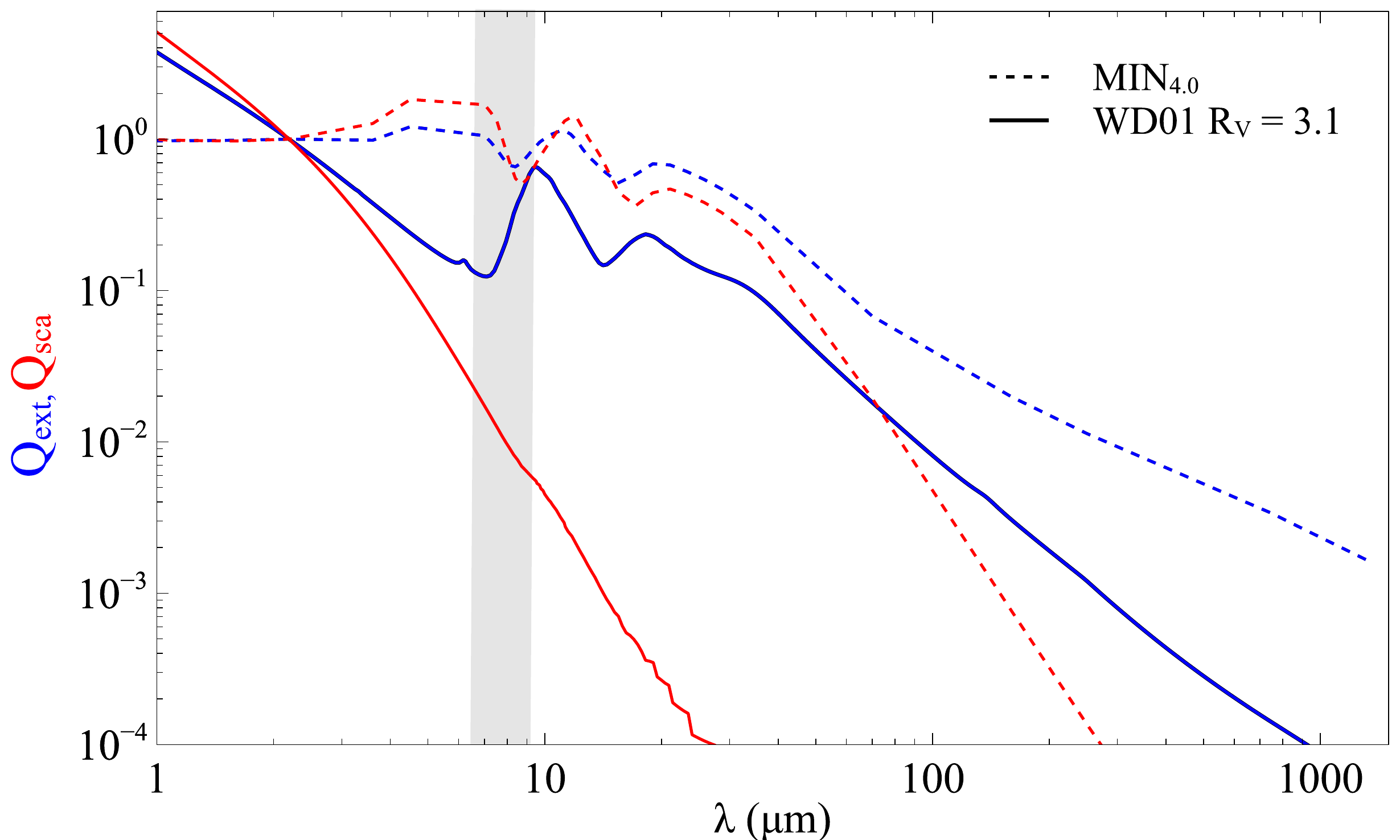}
                \caption{{Dust optical properties (scattering efficiency, Q$_\mathrm{sca}$, in red, total extinction efficiency, Q$_\mathrm{ext}$, in blue). The WD01 (3.1) model is displayed by solid lines, and MIN fluffy aggregates with an equivalent spherical size of 4.0\,\micb are plotted with dotted lines. A gray area represents the 8\,\micb filter width.} All the dust coefficients have been normalized by their value at the K$_\mathrm{s}$ wavelength. }
                \label{fig:dust_model}
        \end{figure}

        Assuming this cloud model, we estimated $\tau_8$ and its associated peak scattering intensity from I$_\mathrm{sca}$ = $\Delta_\mathrm{min}- \mathrm{I}_\mathrm{bg}(\mathrm{e}^{-\tau_\mathrm{8,max}}-1)$. The strength of the scattering component has to vary from $\sim$100\,\kjy\,for the lowest \Ibg value to several hundred \kjy\,for the highest \Ibg value. {With this rough estimate, we expect at least as much} scattering in the 8\,\micb band {as} in the 3.6\,\micb band and, depending on the true background value, possibly more. Among the previous dust models tested in \cite{2014Lefevre}, only a few are able to scatter enough in the 8\,\micb \textit{Spitzer} band. We chose to test two sets of dust models {(Fig. \ref{fig:dust_model})}: the compact spherical dust models from WD01, and {the} fluffy aggregates of monomers {(\citealt{Min2015submitted}, App. \ref{app:min_dust})}. {As a simplification, we chose a bimodal distribution} with one population of small compact spherical grains (WD01, R$_\mathrm{V}$ = 3.1) that dominates the external regions {with no scattering}, and large aggegates (MIN$_{4.0}$) to {progressively fill} the inner region. Using the CRT radiative transfer code of \citet{2003Juvela}, the appropriate balance between the two dust populations was constrained by the coreshine intensity ($\Delta_{3.6}$) with the \Ibg value at 3.6\,\micb taken from \cite{2014Lefevre}.

        \begin{figure*}[t!]
                \hspace{1.4cm}  
                \begin{minipage}{10cm}
                        \includegraphics[scale=0.4]{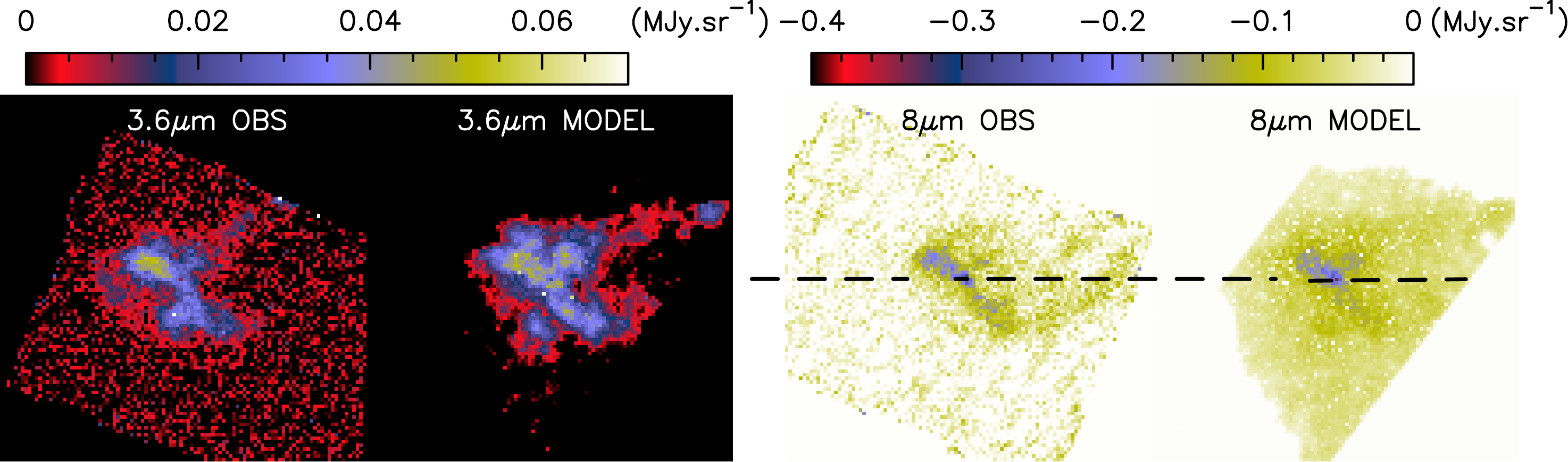}
                \end{minipage}
                \hspace{1.8cm}
                \begin{minipage}{15cm}
                        \includegraphics[scale=0.4]{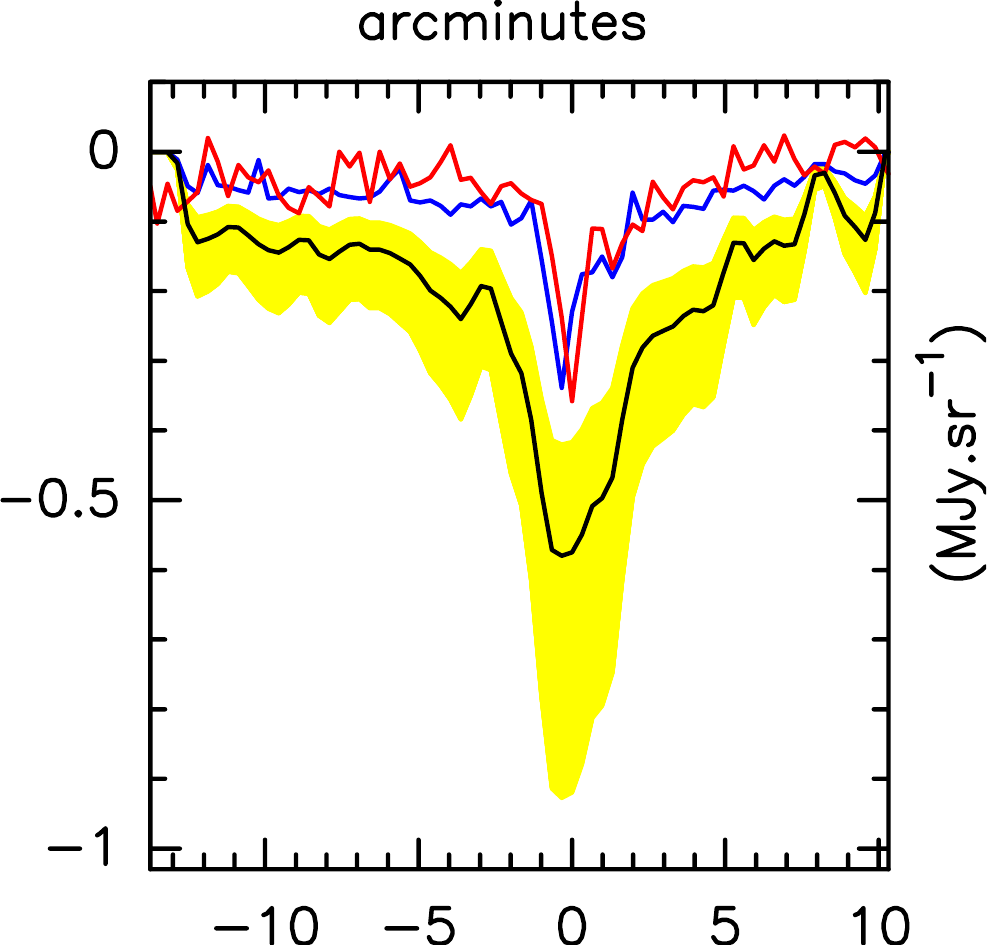}
                \end{minipage}          
                \caption{{From left to right: 3.6\,\micb observations with subtracted background and stars ($\Delta_{3.6}$) compared to its modeling, and the same at 8\,\micb ($\Delta_{8}$). The black dashed line shows the horizontal cut through the core represented in the last panel.} The red line corresponds to the profile from observations ($\Delta_{8}$), and the yellow filled shape displays modeling profiles without I$_\mathrm{sca}$ for \Ibg values ranging from 0.42 to 0.93\,\mjy. The black profile is the one example without I$_\mathrm{sca}$ for \Ibg = 0.58\,\mjy, and the blue one is the same model when including I$_\mathrm{sca}$.}
                \label{fig:profile}
        \end{figure*}

                
        \vspace{-1.0cm}
        
        \section{Compability of the modeling with observations}\label{sect:sca_dust}
        
        Assuming a cloud model compatible with both the NIR extinction and the \ndhp density, we find that the associated $\Delta$ values derived from the models are always deeper than the observed values when not considering I$_\mathrm{sca}$ (Fig. \ref{fig:profile}). {However,} when the I$_\mathrm{sca}$ contribution is taken into account, it is possible to find a suitable solution by including the right abundance of {dust that is} able to scatter at 8\,\mic. Given the degeneracy of the solutions and the assumptions {regarding} the cloud geometry, we chose to illustrate the I$_\mathrm{sca}$ contribution with our bimodal dust populations and focus on their relative abundances. A variation in the relative abundance of the MIN$_{4.0}$ population proportional {to the density} n$^{0.25}$ gives {an adequate match to} the horizontal cut toward the core, as well {as good agreement between the observed and model $\Delta$ values at 3.6 and 8\,\micb}(Fig. \ref{fig:profile}). The dust mass is fixed in our modeling, {which does not include ices}, {and} corresponds to grain growth by coagulation with the density. {Ice coatings may promote coagulation by changing the sticking coefficient of the grains and the scattering efficiency at coreshine wavelength \citep{2014Andersen,2014Lefevre}. However, no ice-coated dust distribution, able to produce efficient scattering at 8\,\micb, is publicly available presently.}

        \begin{table}[t!]
                \centering
                \vspace{-0.4cm}
                \caption{Background values derived from modeling. \Ibg: calculated with $\Delta$ values from Table \ref{tab:stars}. {\Avb: deduced from E(J--K).}}\label{tab:stars_mod}
                \begin{tabular}{ccccccc}
                        \hline Star & $\tau_8^{\ast}$& $\tau_{9.7}^{\ast}$ & I$_\mathrm{sca}^{\ast}$ & \Ibg & \Av \\ 
                        &  & {peak} & \kjy & \mjy& mag\\
                        \hline
                        B & 0.16 $\pm$ 0.02& 0.19  $\pm$ 0.05 &  41 $\pm$ 4& 0.58--0.88 & 2.2 \\ 
                        D & 0.21 $\pm$ 0.02& 0.25  $\pm$ 0.05 & 65 $\pm$ 7& 0.56--0.81 & 2.5 \\                  
                        I & 0.62 $\pm$ 0.06 & 0.47  $\pm$ 0.09 & 170 $\pm$  17& 0.46--0.64 & 8.8\\                 
                        \hline 
                \end{tabular} 
                \vspace{-0.4cm}        
        \end{table}

        {We find that even if the scattering is negligible in front of the stars compared to the stellar flux, {it} is not the case in their vicinity, where the \Ibg value has been deduced.} Taking the I$_\mathrm{sca}$ values in the region of the stars into account, {we calculated again \Ibg\,(see Sect. 3) based on the $\tau_8^{\ast}$ values around Stars B, D, and I obtained from modeling (Table \ref{tab:stars_mod}).} We confirm that the {\Ibg value used for the modeling is now compatible} with the three stars (Fig. \ref{fig:profile}: \Ibg = 0.58\,\mjy). {Moreover,} $\tau_8^{\ast}$ values are similar to the ones calculated with WD01 5.5B (Table \ref{tab:stars}). We also obtained (Table \ref{tab:stars_mod} and App. \ref{app:tau}) compatible values at the silicate absorption peak ($\tau_{9.7}^{\ast}$) with the {ones} of \citet{2013Whittet}. However, grain growth by coagulation is expected to {contribute} to the {relative decrease} in $\tau_{9.7}$ {with respect to} the total dust opacities \citep{2007Chiar}, hence with $\tau_{8}^{\ast}$. {Since grains are coagulating in L183, we expect to observe that $\tau_{9.7}$ would no longer be correlated beyond a given $\tau_8$ threshold, which is what we find from the modeling (App. \ref{app:tau}). This result confirms that $\tau_{9.7}$ might not be a good tracer of the total column density {toward dark clouds}.}
        
        \vspace{-0.5cm}
        \section{Conclusions}\label{sect:conclu}

        From all the validation tests, we {found} that including a fraction of efficiently scattering dust grains (here the MIN$_\mathrm{4.0}$ or any kind of aggregates with a scattering behavior similar to the {red dashed line}, Fig. \ref{fig:dust_model}) is mandatory for reproducing the 8\,\micb observations. Nevertheless, only a few available dust models are efficient enough to produce significant scattering up to 8\,\mic. The WD01 5.5B dust grains are somewhat efficient (I$_\mathrm{sca}\sim$0.1 \mjy), but cannot reproduce the observations for an \Ibg value greater than 0.4 \mjy. Aggregates are much more efficient {at scattering} the light at MIR wavelengths than compact spherical grains, and probably the only way to explain deeper $\Delta$ values observed toward other dense cores (i.e., \object{IRAS16293}, Pagani et al. in prep.). They are also compatible with grain growth by coagulation and with the expected deviation from the correlation of $\tau_{9.7}$ and $\tau_8$ with $\mathrm{E(J-K)}$. This is also {in accord} with the idea that grains remain fluffy even in the dense prestellar cores and protoplanetary disks \citep{2013Mulders}.\\
        We demonstrated the importance of scattering up to 8\,\micb and the necessity of using realistic dust models. The final  $\tau_8$  map obtained from the modeling is compatible with a filament  at \Av$>${25\,mag} and a central density for the core of 1.8$\times$10$^6$ \cc.  Multiwavelength modeling, including NIR wavelengths, will {place} more constraints on the cloud shape and small grains, as discussed in \cite{2014Lefevre}. {Since ices are present at relatively low extinction and can have an impact on coagulation and scattering efficiencies, smaller aggregates with ices \citep{2015Kohler} will be investigated in a future paper}. Adopting such dust models will also have consequencies on the interpretation of data in emission since {aggregates are known to be better emitters \citep{2003Stepnik}. In this paper, the emissivities of the aggregates are higher by at least one order of magnitude at far-infrared wavelengths than for spherical grains (Fig. \ref{fig:dust_model}).}

        \begin{acknowledgements}
                \thanks{This work was supported by the CNRS program "Physique et Chimie du Milieu Interstellaire" (PCMI), the DIM ACAV and "RÈgion Ile de France". CP and DW acknowledge support from the NASA Astrobiology and Exobiology programs. We thank N. Ysard for fruitful discussions.}
        \end{acknowledgements}

        \Online
        \begin{appendix}
                
                \section{Regions with significant scattering}\label{app:discr}
                
\begin{figure}[h!]
                                        \centering
                                        \includegraphics[width=0.8\linewidth]{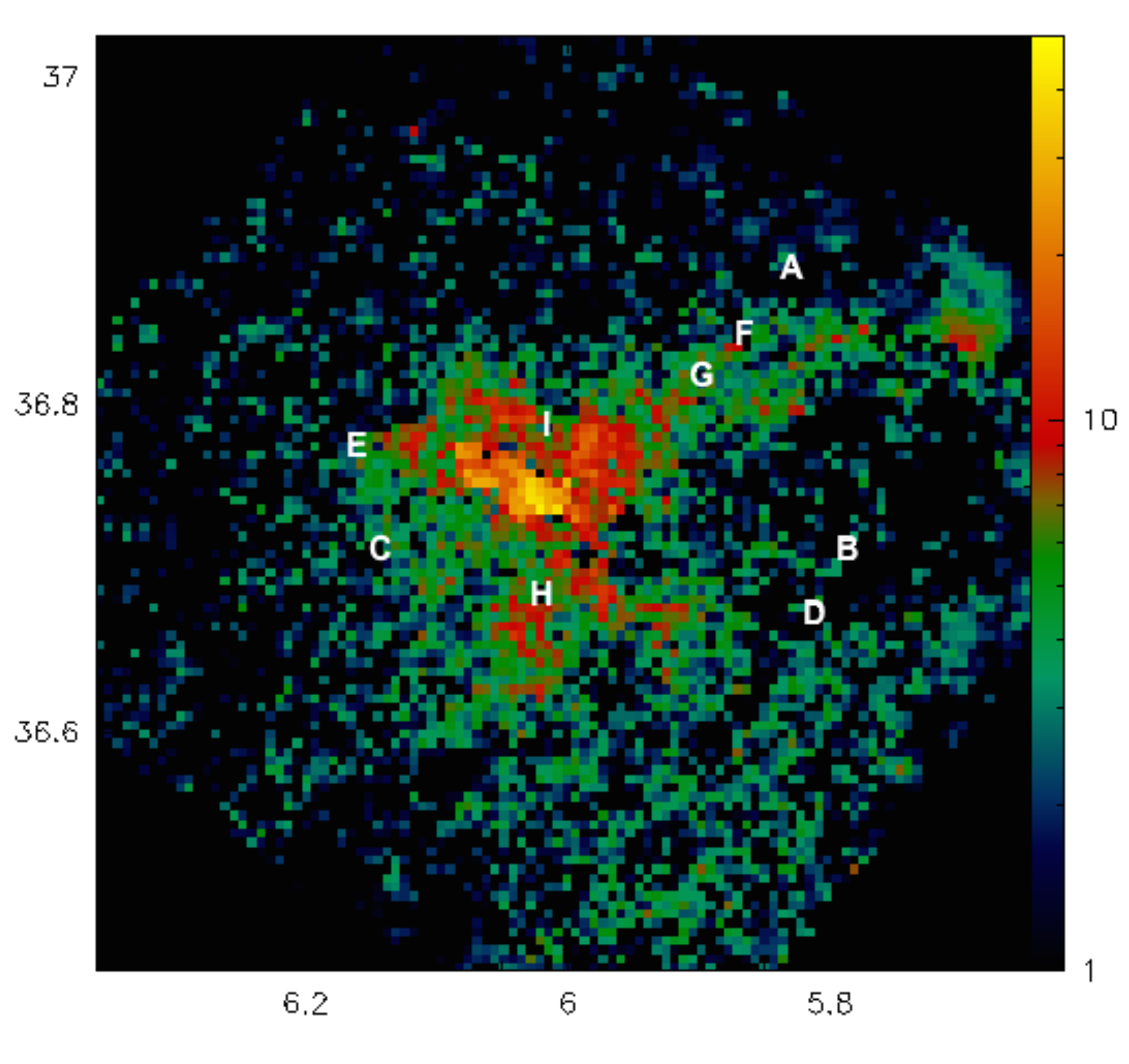}
                                        \caption{Difference between the visual extinction obtained via the reddening and the one via the 8\,\micb map inversion (in magnitudes). The letters refer to the stars studied by \citet{2013Whittet}.}
                                        \label{fig:spatial_discr}
                                \end{figure}

                {The discrepancy between the 8\,\micb map inversion and other estimates for the visual extinction (Fig. \ref{fig:Av_converted}) is not a simple offset that could be applied to the whole map, but it varies throughout the cloud (Figure \ref{fig:spatial_discr}). While green regions in the outer part can be interpreted as a lack of information in the 8\,\micb map due to the noise, other colors trace regions where the scattering cannot be neglected ($\gtrsim$ 25 mag).}
                \vspace{-0.6cm}
                
                \vspace{-0.8cm}
                
                \section{The optical properties of aggregate particles}\label{app:min_dust}
                
                The optical properties of dust particles are very sensitive to their shape. In particular, perfect spherical particles represent a class of particles that are incompatible with observations and laboratory measurements. Therefore, we consider it very important for detailed studies of scattering properties of dust grains to use a realistic dust particle model. All details on the construction of the fluffy aggregates and their optical properties will be presented in \cite{Min2015submitted}. Below we give the basic properties needed for a proper understanding of the results presented in this paper.
                
                The optical properties of the aggregate particles are computed using the so-called Discrete Dipole Approximation \citep[DDA;][]{1973ApJ...186..705P, 1994JOSAA..11.1491D}. With this method one can compute the optical properties of particles with arbitrary shape and composition. Despite the confusing term "approximation" in the name of this method, it is actually an exact method in the limit of infinite spatial resolution. We construct monomers of the aggregates by using Gaussian random field particles (Min et al. 2007). We then glue these monomers together to construct fluffy aggregates. In this way we avoid all effects of sphericity of the particles since the monomers of the aggregates are also non-spherical.
                
                Each monomer in the aggregate is made of a single material. We randomly assign a material to each monomer using the overall composition: 75\% silicate, 15\% carbon, and 10\% iron sulfide (by volume). This composition is roughly consistent with the solar system composition proposed by \citet{2011Icar..212..416M}. We use the refractive index data from \citet{1995A&A...300..503D}, \citet{1993A&A...279..577P}, and \citet{1994ApJ...423L..71B} for the silicate (MgSiO$_3$), carbon, and iron-sulfide particles, respectively.
                
                Using the DDA code ADDA \citep{2011JQSRT.112.2234Y}, we compute the absorption, scattering, and extinction cross sections, as well as the full scattering matrix elements at wavelengths from the visible up to millimeter wavelengths for aggregates composed of 1 to 8000 monomers (corresponding to volume equivalent radii from 0.2 to 4.0$\,\mu$m).

                \section{}\label{app:tau}
                
\begin{figure}[h!]
                                        \centering
                                        \includegraphics[width=0.85\linewidth]{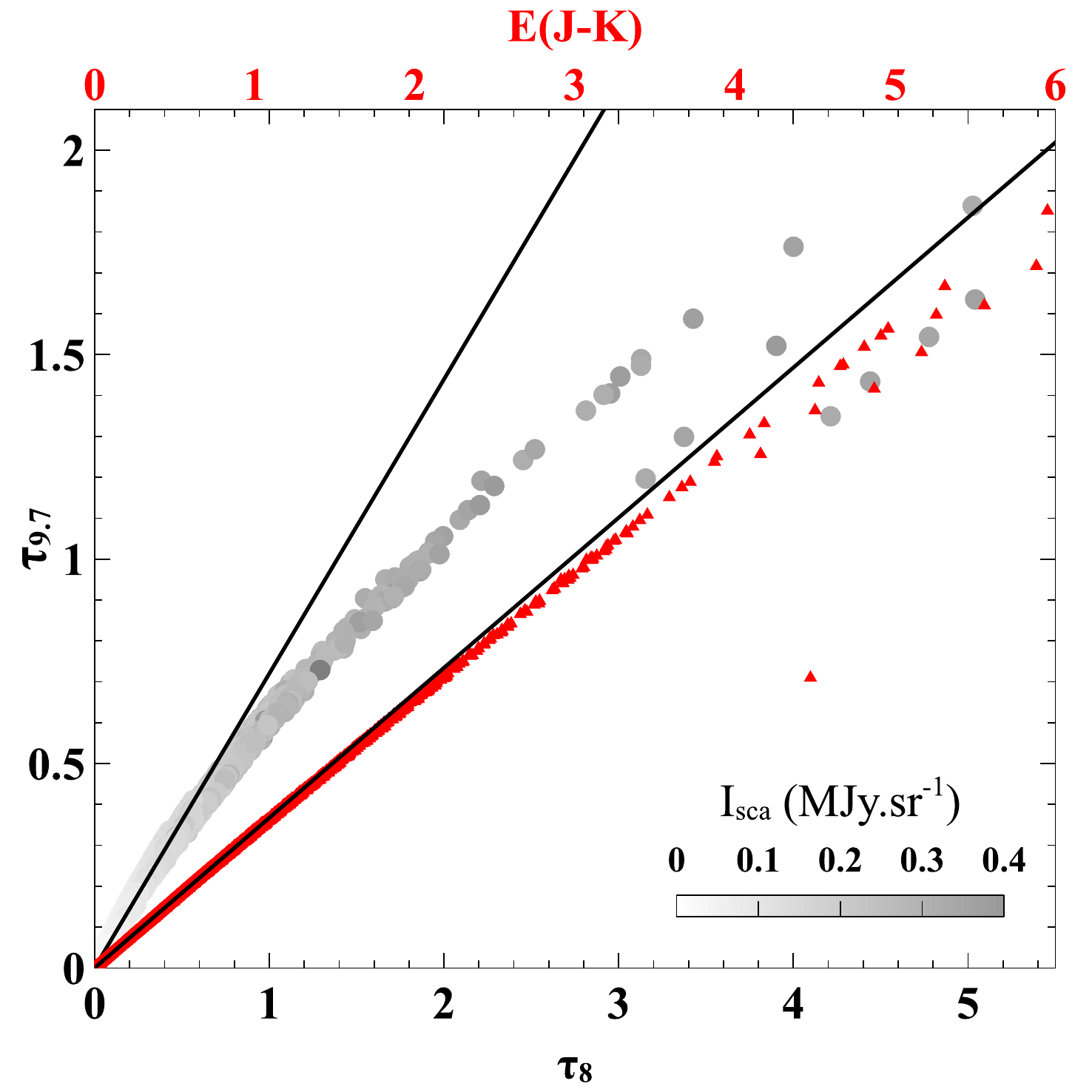}
                                        \caption{The 9.7\,\micb silicate absorption line opacities as a function of $\tau_8$ ({gray circles}). The color of the {circles} is related to the I$_\mathrm{sca}$ strength. {The 9.7\,\micb silicate absorption line opacities as a function of E(J--K) is represented by {red} triangles.} All the values are obtained from the modeling {and {black} lines correspond to the correlation obtained from diffuse interstellar lines of sight {\citep{2003Whittet}}.}}
                                        \label{fig:EKK8}
                                \end{figure}

                {To compare the $\tau_{9.7}$ values measured at the silicate absorption peak by \cite{2013Whittet} with the one from the modeling, $\tau_{9.7}^{\ast}$, we subtracted the continuum opacity from the total dust opacity at 9.7\,\mic.} In order to verify the conversion between $\tau_8$ and $\tau_{9.7}$, we made a plot of the values cell by cell (Fig. \ref{fig:EKK8}). 
                
                For $\tau_8$ lower than 0.5, both are correlated by the relation $\tau_{9.7}$/$\tau_8$ $\sim$ 0.72 (red line Fig. \ref{fig:EKK8}) but above this threshold, the $\tau_{9.7}$ values decrease more rapidly than the correlation law. The behavior of $\tau_{9.7}$ with $\tau_{8}$  is also likely to be part of the explanation of the effect observed between $\tau_{9.7}$ and E(J--K) by \cite{2007Chiar} and by \citet[towards L183]{2013Whittet}. Indeed, beyond a limit of $\mathrm{E(J-K)}=2\,\mathrm{mag}$, the measured optical depth of the 9.7\,\micb silicate absorption feature ($\tau_{9.7}$) {also decreases with respect to {E(J--K) measured} along diffuse interstellar lines of sight.} Since $\tau_{9.7}$ {measures the silicate absorption along the line of sight and not the continuum opacity from other dust species, it may not be representative of the total dust opacity toward dark clouds.} Moreover, if the line of sight opacity is dominated by small silicate grains, {the $\tau_{9.7}$ is expected to be less than proportional to $\tau_8$ with coagulation.} We detect this effect from our modeling (Fig. \ref{fig:EKK8}) associated to an increase in I$_\mathrm{sca}$ that traces the coagulation. Since J and K bands are very sensitive to the cloud model \citep{2014Lefevre},  {the $\tau_{9.7}$ function of E(J--K), plotted in Fig. \ref{fig:EKK8}, must be considered as a simple trend at this stage.}

                \section{Institutional and technical acknowledgements}
                This research has made use of observations from the Spitzer Space Telescope and data from the NASA/IPAC Infrared Science Archive, which are operated by the Jet Propulsion Laboratory (JPL) and the California Institute of Technology under contract with NASA. It makes use of data products from the Wide-field Infrared Survey Explorer, which is a joint project of the University of California, Los Angeles, and the Jet Propulsion Laboratory/California Institute of Technology, funded by the National Aeronautics and Space Administration. It makes use of data products from the Two Micron All Sky Survey, which is a joint project of the University of Massachusetts and the Infrared Processing and Analysis Center/California Institute of Technology, funded by the National Aeronautics and Space Administration and the National Science Foundation. Our VISTA data are based on observations made with ESO Telescopes at the  Paranal Observatory under program ID 091.C-0795. We thank L. Cambr{\'e}sy and M. Juvela for providing their code to deduce NIR extinction, and M.J. for CRT. We thank P. Hudelot and N. Bouflous for the TERAPIX data treatment {of our VISTA data}.
                
        \end{appendix}

\end{document}